\documentclass[twocolumn,superscriptaddress,preprintnumbers,amsmath,amssymb,prb]{revtex4-2}

\newcommand{\IMSS}{Muon Science Laboratory and Condensed Matter Research Center, Institute of Materials Structure Science, High Energy Accelerator Research Organization (KEK-IMSS), Tsukuba, Ibaraki 305-0801, Japan}
\newcommand{\Sokendai}{Department of Materials Structure Science, The Graduate University for Advanced Studies (Sokendai), Tsukuba, Ibaraki 305-0801, Japan}

\newcommand{\MCES}{Materials Research Center for Element Strategy, Tokyo Institute of Technology (MCES), Yokohama, Kanagawa 226-8503, Japan}

\usepackage{txfonts}
\usepackage[dvipdfmx]{graphicx}
\usepackage{dcolumn} 
\usepackage{bm} 
\usepackage{multirow}
\usepackage[version=3]{mhchem} 
\begin{document}

\title{Anomalous diamagnetism of electride electrons in transition metal silicides}

\author{M.~Hiraishi}
\affiliation{\IMSS}
\author{K.~M.~Kojima}
\affiliation{\IMSS}\affiliation{\Sokendai}
\author{H.~Okabe}
\affiliation{\IMSS}
\author{A.~Koda}
\affiliation{\IMSS}\affiliation{\Sokendai}
\author{R.~Kadono}
\affiliation{\IMSS}\affiliation{\Sokendai}
\author{J.~Wu}\affiliation{\MCES}
\author{Y.~Lu}\affiliation{\MCES}
\author{H.~Hosono}
\affiliation{\MCES}
\begin{abstract}
  Intermetallic silicide compounds, LaScSi and Y$_5$Si$_3$, known for being hydrogen (H) storage materials, are drawing attention as candidates for electrides in which anions are substituted by unbound electrons.
  It is inferred from a muon spin rotation experiment that the local field at the muon site (which is the same site as that for H) in these compounds exhibits a large negative shift under an external magnetic field, which is mostly independent of temperature.
  Such anomalous diamagnetism signals a unique property of electride electrons associated with transition metals.
Moreover, the diamagnetic shift decreases with increasing H content, suggesting that the electride electrons existing coherently in the hollow interstitial positions are adsorbed by H to form hydride ions (H$^-$).
\end{abstract}

\maketitle
\section{INTRODUCTION}
Electrides are a class of materials in which excess electrons are accommodated in voids as anions without atomic nuclei~\cite{Dye2003,Hosono_Kitano_Review2021}. They have been drawing much attention because of their high electron mobility, low work function, and high affinity for hydrogen.
The recent demonstration of oxygen-depleted mayenite ([Ca$_{24}$Al$_{28}$O$_{64}]^{4+}$4e$^-$, abbreviated as C12A7:e$^-$) as the first air-stable solid electride~\cite{c12a7_matsuishi2003,c12a7_matsuishi2008,c12a7_toda} has started extensive research on solid electrides. This has resulted in a breakthrough in ammonia synthesis due to the very high catalytic efficiency of ruthenium (Ru) dispersed on C12A7:e$^-$ (Refs.~\onlinecite{c12a7_kitano,c12a7_kitano2015}), thus leading to various applications in related fields~\cite{c12a7_kim,c12a7_bucha,c12a7_ruszak,c12a7_adachi,c12a7_toda2013}.
A search for candidate solid electride compounds has also been triggered. Intermetallic silicides, which include LaScSi and Y$_5$Si$_3$, have been proposed as a new subclass of electrides~\cite{WuAdvMat17,Y5Si3}.
These compounds are stable under atmospheric as well as moist conditions and have received growing interest because they can also serve as highly efficient catalysts for ammonia synthesis when dispersed with Ru.
\par
\begin{figure}[t]
  \centering
	\includegraphics[width=\linewidth]{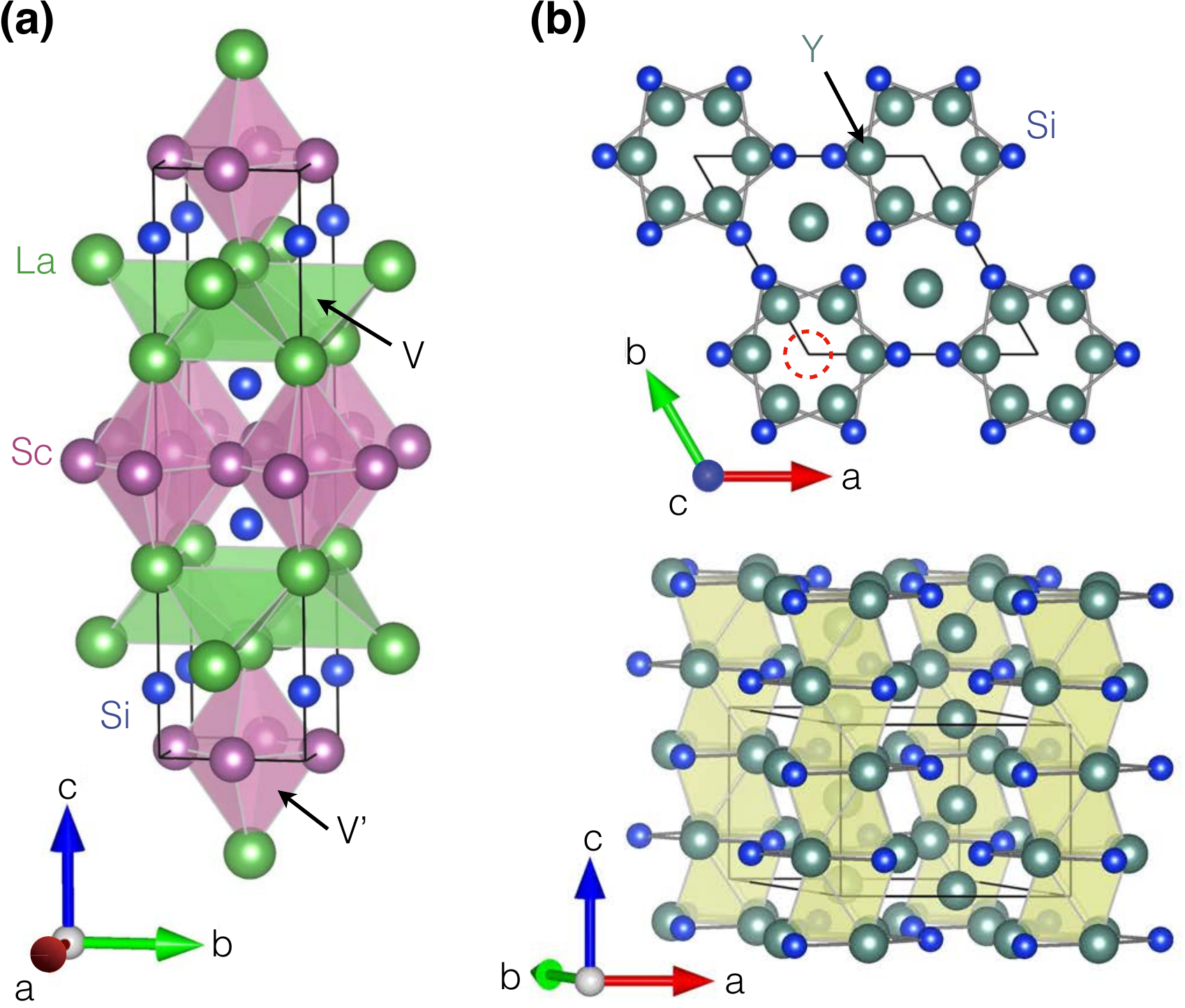}
  \caption{Crystal structures of LaScSi (a) and Y$_5$Si$_3$ (b). The red dashed circle in the upper part of (b) indicates a one-dimensional channel (hole) running along the $c$ axis.}
  \label{Fig0}
\end{figure}

LaScSi (space group $I4/mmm$) has layered networks of La$_4$ tetrahedral voids (V) and La$_2$Sc$_4$ octahedral voids (V') interlaced with Si layers [see Fig.~\ref{Fig0}(a)].
X-ray diffraction (XRD) and thermal desorption spectroscopy have shown that up to 1.5 hydrogen atoms can be stored in the V and V' sites per formula unit~\cite{LaScSiH, WuAdvMat17}.
First-principles density functional theory (DFT) calculations have shown that the electrons that contribute to the density of states near the Fermi energy [$N(E_\mathrm{F})$] are localized at the center of the V and V' sites, depending on the H concentration.
Because the electride electrons in pristine LaScSi are predicted to be mostly localized at the V sites, a stepwise hydrogenation process has been proposed. In this process, H first preferentially occupies the V site to form LaScSiHV'$_{0.5}$ and then the V' site to form LaScSiH$_{1.5}$ (tiered electron anions)~\cite{WuAdvMat17}.
\par
Y$_5$Si$_3$ belongs to the Mn$_5$Si$_3$-type structure [space group $P6_3/mcm$, see Fig.~\ref{Fig0}(b)] and has been studied mainly as a hydrogen storage alloy~\cite{Y5Si3H}.
The hydride, Y$_5$Si$_3$H$_x$, has been reported to accommodate H in a position coordinated by six Y atoms (2$b$ site) that comprises a one-dimensional hollow channel of approximately 0.4~nm diameter running along the $c$-axis.
DFT calculations have indicated that a band originating from the $1s$ orbital of H exists at approximately 5~eV below $E_\mathrm{F}$ in Y$_5$Si$_3$H, whereas another band strongly hybridized with the 4$d$ orbital of Y exists near $E_\mathrm{F}$ in pristine Y$_5$Si$_3$~\cite{Y5Si3}.
The latter is presumed to originate from electrons localized at the $2b$ site (electride electrons) and contributes to the emergence of functionalities such as catalytic activity.

\par
In C12A7:e$^-$, hydride ions are localized at the center of the cage where electride electrons reside upon H intake~\cite{c12a7_matsuishi2003,C12A7H}.
Because of the differences in the local structure of the electride electrons and their band characteristics between LaScSi/Y5Si and C12A7:e$^-$ (anionic electron bands strongly overlap with other atomic orbitals in the former~\cite{WuAdvMat17,Y5Si3} and the bands are virtually isolated in the latter~\cite{c12a7_matsuishi2008}), it is important to understand the local electronic state associated with H occupancy. This state plays an important role in catalytic activity~\cite{c12a7_kitano,c12a7_kitano2015}.
\par
In this regard, it is worth mentioning that a peculiar diamagnetism has been reported in a muon spin rotation ($\mu$SR) study of Y$_2$C (Ref.~\onlinecite{Y2C_muon}), which is also known to be a two-dimensional electride with H-storage capability~\cite{Y2C_DFT2019,Inoshita_Y2C,Zhang_Y2C}. Because muons acting as pseudo-H occupy the H site, which is situated in the center of the electride electrons spread over two dimensions, it has been argued that the diamagnetism probed by muons is characteristic to the band structure of electrides that exhibit strong hybridization with Y 4$d$ orbitals. Theoretical prediction of ferromagnetic instability (Stoner ferromagnetism) and inelastic neutron scattering measurement in Y$_2$C suggest a possible link between the electride electrons and Y 4$d$ bands~\cite{Inoshita_Y2C,y2c_tamatsukuri}.
\par
The implanted muon can be regarded as a light isotope of H (approximately 1/9$m_\mathrm{H}$, where $m_\mathrm{H}$ is the mass of hydrogen), which provides a unique tool for experimentally simulating interstitial H in the dilute limit. This allows spectroscopic information on the electronic structure to be obtained via $\mu$SR. In particular, $\mu$SR can provide information on the local valency of muons as pseudo-H (corresponding to H$^\pm$, H$^0$), which is not readily accessible by other conventional techniques.
In contrast to the large difference in the mass between a muon and a proton, the difference between the reduced mass of the bound electrons is only approximately 0.4\%. Thus, the local electronic structure associated with muons is almost identical to that of protons in the host matter. As the term ``muonium" refers exclusively to the neutral bound state of $\mu^+$ and $e^-$ (i.e., to the {\sl structure}), it would be convenient to introduce the term ``muogen'' (abbreviated as Mu) to describe the {\sl properties} of muons as an element ($\ce{_{1}^{0.1126}Mu}$)~\cite{FeS2_Okabe,IGZO_muon}.
\par

Here, we report on the microscopic properties of H and electride electrons probed by $\mu$SR in LaScSi and its hydrides and in Y$_5$Si$_3$. In LaScSi and its hydrides/deuterides, the Mu occupancy of V/V' sites as pseudo-H was determined through a detailed analysis of the $\mu$SR spectra under zero and longitudinal fields. The results were confirmed to be consistent with those inferred from DFT calculations for H at the V and V' sites in the dilute limit. The interplay between H and the electride electrons
in these compounds was investigated through the $\mu$SR frequency shift under a magnetic field, which is proportional to the local susceptibility at the Mu site. An anomalously large negative shift was observed in pristine compounds irrespective of temperature, for which we argue a couple of possible mechanisms.
The shift decreased monotonously with increasing H content, suggesting that the electride electrons existed in a state delocalized over multiple H-sites and that they were localized by H to form hydride ions in the voids.

\section{EXPERIMENTAL METHODS and DFT CALCULATIONS}\label{SecII}
The polycrystalline samples of LaScSi and Y$_5$Si$_3$ used in the $\mu$SR experiment were synthesized using the arc-melting method.
The details of the sample synthesis can be found in the literature~\cite{Y5Si3, WuAdvMat17}.
For LaScSi, hydrogenated and deuterated samples were prepared to investigate the mutual influence between the Mu (as pseudo-H) and electride electrons.
The four samples comprise two samples of LaScSiH$_x$ ($x=0.9$-1.2 and $x\sim1.5$, which are denoted as $x=1$ and $x=1.5$, respectively) and two samples of LaScSiD$_x$ ($x=0.4$-0.5 and $x=1.2$-1.5, which are denoted as $x=0.5$ and $x=1.2$, respectively).
We note that there was some uncertainty in the D content of the deuterated sample because the D content was determined empirically by comparing the results of the D content analysis with those obtained earlier for LaScSiH$_x$ (without the standard reference for D).
Additional details for experiments and DFT calculations are provided in the Supplemental Material~\cite{supplement}.
\section{RESULTS}
\subsection{Mu sites: LaScSiH$_x$}
\begin{figure}[bt]
  \centering
	\includegraphics[width=\linewidth]{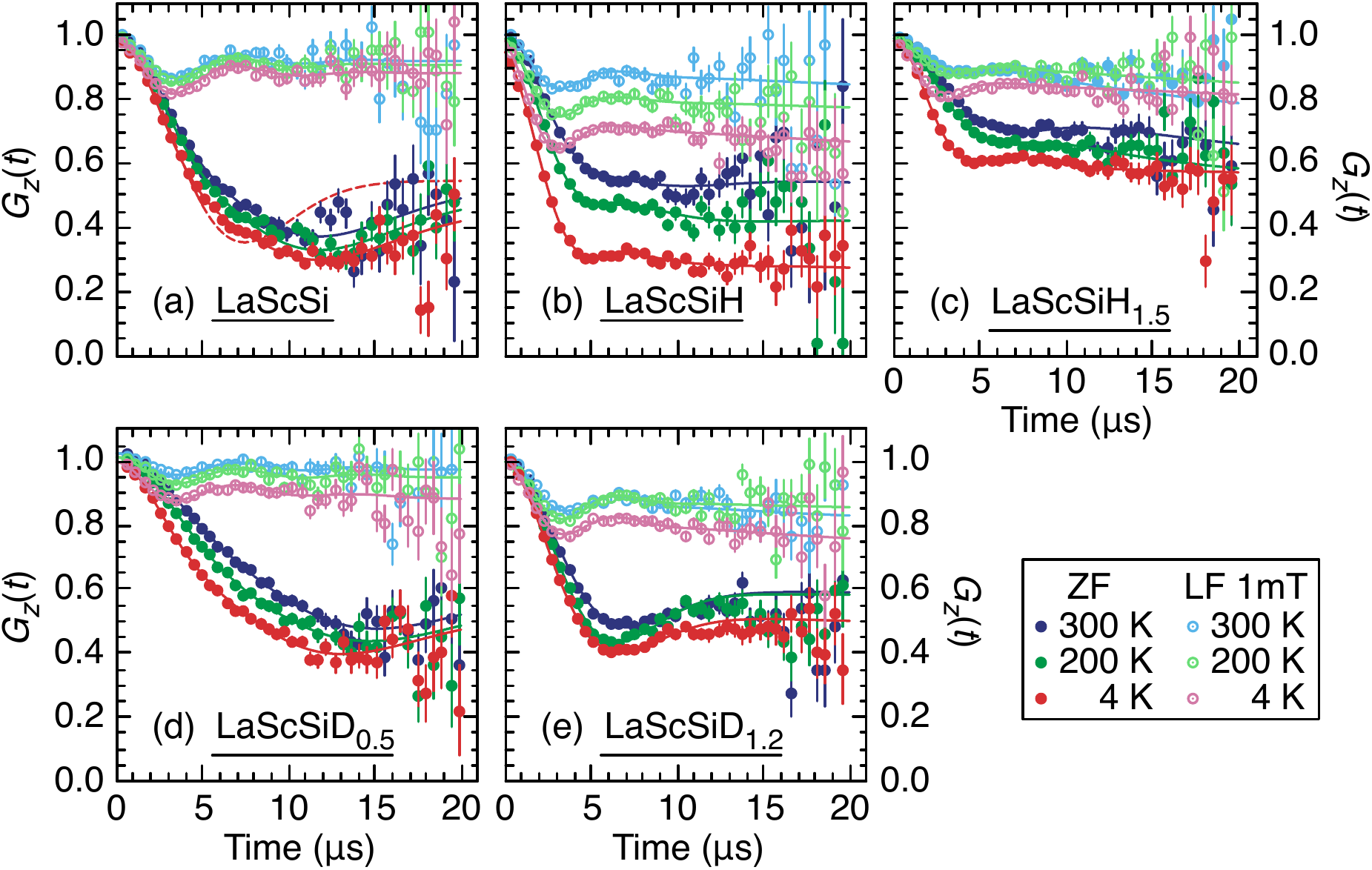}
  \caption{$\mu$SR time spectra under zero (ZF, closed symbol) and longitudinal field (LF=1~mT, open symbol) observed in LaScSiH$_x$ with $x=0$ (a), $x=1$ (b), and $x=1.5$ (c) and in LaScSiD$_x$ with $x=0.5$ (d) and $x=1.2$ (e). The solid and dashed curves in (a) are the results of curve-fits using Eq.~(\ref{Eq1}) with $n=1$ and $n = 2$, respectively.}
  \label{Fig2}
\end{figure}

Figure~\ref{Fig2}(a) shows the $\mu$SR time spectra observed in LaScSi under zero external field (ZF) and a longitudinal field (LF) of 1~mT. The spectra are normalized by the initial asymmetry [$A(0)=A_0$] observed at ambient temperature.
The ZF-$\mu$SR spectrum exhibits a slow Gaussian depolarization due to the random local fields from the nuclear magnetic moments. $A_0$ is largely independent of temperature.
These features indicate that most of the Mu is in a diamagnetic state (Mu$^+$ or Mu$^-$).
As some of the incident muons are stopped at the backing material (silver), which has negligible depolarization, the time spectra are analyzed by using the following function:
\begin{align}
  A(t)=A_0G_z(t)=\sum_{i=1}^nA_iG_\mathrm{KT}(\Delta_i,B_\mathrm{ext},\nu_i,t)+A_\mathrm{c},
  \label{Eq1}
 \end{align}
 which is approximated as
\begin{align}
  A(t) \simeq\sum_{i=1}^nA_i\left[\frac{1}{3}e^{-\nu_it}+\frac{2}{3}(1-\Delta_i^2t^2)e^{-\Delta_i^2t^2/2}\right]+A_\mathrm{c},\nonumber
\end{align}
for the case of $\nu_i\ll\Delta_i$ and ZF ($B_\mathrm{ext}=0$). $A_i$ is the initial asymmetry of the Mu occupying the $i$-th site described by the Gaussian Kubo--Toyabe function, $G_\mathrm{KT}(\Delta_i,B_\mathrm{ext},\nu_i,t)$, where $\Delta_i$ denotes the linewidth determined by the RMS of the corresponding local field distribution (see Supplemental Material~\cite{supplement}), $\nu_i$ is the fluctuation rate of $\Delta_i$, and $B_\mathrm{ext}$ is the external LF~\cite{Hayano}.
The constant term, $A_\mathrm{c}$, consists of two contributions: one from the Mu stopped in the backing and another from the Mu that is present in the sample but exhibits no depolarization (see below).

In Fig.~\ref{Fig2}(a), the results of the least-squares fits of the ZF-$\mu$SR spectra using Eq.~(\ref{Eq1}) are shown for $n=1$ (red dashed curves) and 2 (solid curves). The latter shows a significantly better agreement with the experimental results.
This indicates that the Mu probes different magnetic field distributions at two crystallographically inequivalent sites in LaScSi.
The parameter values deduced from the curve-fit at the lowest temperature (4~K) are $\Delta_1=0.1468(33)~\mu$s$^{-1}$ and $\Delta_2=0.347(6)~\mu$s$^{-1}$.
The calculated $\Delta$ values at the V and V' sites using Eq.~(S1) in Supplemental Materials~\cite{supplement} and the lattice structure obtained from the DFT calculation [in which structural relaxation was allowed for one H in a $4\times4\times1$ supercell (192 atoms) to mimic the isolated Mu] are 0.1465 and 0.313~$\mu$s$^{-1}$, which are in reasonable agreement with $\Delta_1$ and $\Delta_2$, respectively (see Table~\ref{table}).
Thus, we attribute $A_1$ ($A_2$) to the signals from the Mu at the V (V') sites.
\par

\begin{table*}
	\centering
	\caption{Linewidth $\Delta_i$ in LaScSiH/D$_x$ deduced from ZF/LF-$\mu$SR measurements at 4~K. The simulated values for V/V' sites in LaScSi correspond to the relaxed lattice structure obtained from the DFT calculations in the $4\times4\times1$ superlattice (192 atoms) with one hydrogen at the center of the V/V' site, whereas the others are based on structural parameters obtained by XRD~\cite{WuAdvMat17}.}
  \renewcommand\arraystretch{1.3}
  \tabcolsep=2mm
  \begin{tabular}{ccccccc}
		\hline\hline
    & & LaScSi & LaScSiH & LaScSiH$_{1.5}$ & LaScSiD$_{0.5}$ & LaScSiD$_{1.2}$\\
    \multirow{2}{*}{Experiment} & $\Delta_1~(\mu$s$^{-1})$ & 0.1468(33) & 0.183(7) & 0.174(9) & 0.135(4) & \multirow{2}{*}{0.286(1)}\\
    & $\Delta_2~(\mu$s$^{-1})$ & 0.347(6) & 0.415(4) & 0.431(4) & 0.348(9) & \\\hline
    \multirow{2}{*}{Simulation} & V center~$(\mu$s$^{-1})$ & 0.1465 & 0.191 & 0.194 & 0.138 & 0.144\\
		& V' center~$(\mu$s$^{-1})$& 0.313 & 0.342 & 0.345 & 0.338 & 0.340
    \\\hline \hline
 	\end{tabular}
	\label{table}
\end{table*}

As shown in Fig.~\ref{Fig2}(b,c,d,e), the depolarization rates ($\propto\Delta_i$) in LaScSiH$_x$ and LaScSiD$_x$ are enhanced with increasing H/D content. This enhancement is explained by the preferential occupancy of the V' sites (where the Mu exhibits a greater depolarization rate) when the V sites are fully occupied by H/D.
Moreover, $A_c$ exhibits an increase with $x$, as well as temperature. (see Supplemental Material~\cite{supplement}, Fig.~S1.)
Although the increase in $\Delta_i$ can be qualitatively attributed to the additional contribution from the nuclear magnetic moments of the nearby H, understanding the latter behavior is not trivial. As no interstitial sites are free of local magnetic fields from $^{139}$La and $^{45}$Sc nuclei, which have 100\% natural abundance, the increase in the undamped component ($\Delta\sim$~0) suggests that a part of the implanted Mu is subject to fast spin dynamics of unknown origin. It should be noted that a similar behavior has been reported for the hydrogen storage material, NaAlH$_4$~\cite{NaAlH4}.
An increase in $A_\mathrm{c}$ above approximately 100~K is observed for all the compositions studied, and the increase is the most prominent in LaScSiH. Elucidation of the origin of these behaviors is an open issue to be addressed in the future.
\par
With the exception of LaScSiD$_{1.2}$, the curve-fit analysis requires two components ($n=2$) in Eq.~(\ref{Eq1}), suggesting that both V and V' sites are occupied by the Mu, regardless of the H/D content.
The fact that the observed linewidth in LaScSiD$_{1.2}$ [$\Delta=0.286(1)$ MHz] is nearly identical to the mean of those estimated for the two sites (0.144 and 0.340 MHz) suggests a similar scenario in LaScSiD$_{1.2}$.
The dispersed H occupancy over both the V and V' sites can be explained by the gain in entropy. Thus, the vacant V sites are made available to the Mu (see the next paragraph). The observed value of $\Delta_1$ is in reasonable agreement with that calculated for the V site irrespective of the H content; however, $\Delta_2$ is approximately 20\% larger than that estimated for the V' site in LaScSiH and LaScSiH$_{1.5}$ (see Table~\ref{table}).
In NdScSiD$_{1.5}$, which has the same crystal structure as LaScSiH$_{1.5}$, neutron diffraction measurements have indicated that D at the V' site is localized slightly off the $c$-axis from the Sc$_4$ plane~\cite{NdScSiD}.
However, this displacement leads to a decrease in $\Delta$, in contrast to the observed trend.
Therefore, we tentatively attribute the discrepancy to the possibility that the local field distribution differs from the Gaussian distribution (a prerequisite of the Kubo--Toyabe function) because of the random occupation of the V' site by H near the Mu.
\par
\begin{figure}[tb]
  \centering
	\includegraphics[width=0.65\linewidth]{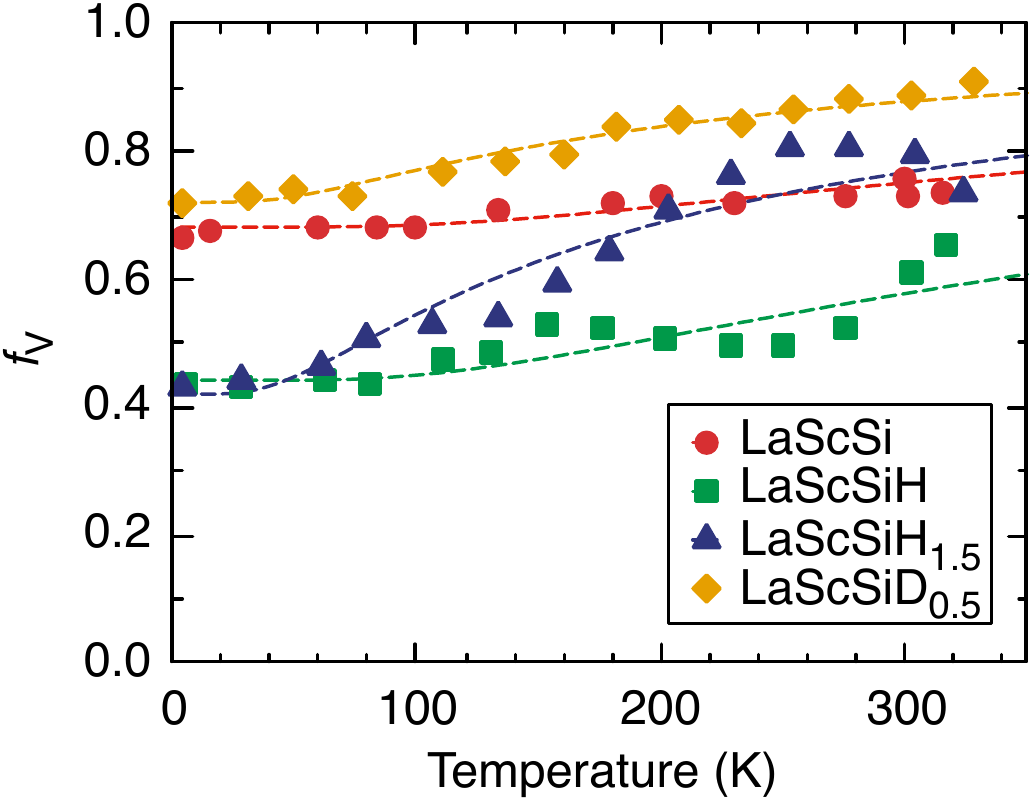}
  \caption{Temperature dependence of fractional yields for muons localized at the V site determined by $f_\mathrm{V}\equiv A_1/(A_1+A_2)$. The errors are smaller than the symbol size. The dashed curves are the best-fits of Eq.~(\ref{Eq_fv}).}
  \label{Fig3}
\end{figure}
As inferred from the temperature dependence of $f_\mathrm{V}$ shown in Fig.~\ref{Fig3}, the occupancy of the V sites increases with temperature regardless of the H/D content, indicating that the Mu preferentially occupies the V sites at higher temperatures.
This is consistent with the general assumption that the initial Mu occupancy is that of a quenched state (corresponding to $T\rightarrow\infty$) determined by the density of the available sites (e.g., $f_\mathrm{V}\simeq2/3$ for $x=0$). As the temperature increases, $f_\mathrm{V}$ is expected to approach the equilibrium distribution [$\simeq2/(2+e^{-\varDelta\varepsilon/k_\mathrm{B}T}$) at $x=0$, where $\varDelta\varepsilon$ is the energy difference between the two sites]. Note that the H occupancy is described by the equilibrium distribution, where H exhibits a smaller V site occupancy at higher temperatures.

Considering that the Mu is a microscopic entity for probing the local motion averaged over a time range limited by the muon lifetime (approximately $10^1$ $\mu$s), we assume that the temperature dependence of $f_\mathrm{V}$ is determined by the migration of the muons from the V' to V sites via hopping motion, which is controlled by a potential barrier, $\varepsilon$ (not $\varDelta\varepsilon$):
\begin{align}
  f_\mathrm{V}(T)=f^0_\mathrm{V}+(1-f^0_\mathrm{V})e^{-\varepsilon/k_\mathrm{B}T},
  \label{Eq_fv}
\end{align}
where $f^0_\mathrm{V}$ is the initial Mu occupancy of the V sites (see Supplemental Material~\cite{supplement}).
The dashed curves in Fig.~\ref{Fig3} are fitted with the aforementioned model. The temperature dependence is reasonably reproduced except for LaScSiH in which $A_\mathrm{c}$ is significantly enhanced at high temperatures. The deduced parameter values are $f_\mathrm{V}^0=0.682(5)$, 0.44(2), 0.41(3), and 0.721(9) and $\varepsilon=39(3)$, 37(4), 13(1), and 15(1)~meV for LaScSi, LaScSiH, LaScSiD$_{0.5}$, and LaScSiH$_{1.5}$, respectively.
These results are consistent with the assumption that the Mu localizes at the V site as the true ground state at higher temperatures; this is also in line with the stepwise hydrogenation process proposed in the literature~\cite{WuAdvMat17}.
Bader charge analysis~\cite{Bader} of H localized at the V and V' site centers in the dilute limit results in H$^{-0.63}$ and H$^{-0.80}$, respectively, suggesting that the corresponding Mu charge state is Mu$^-$.
\subsection{Anomalous diamagnetism of electride electrons}
In the high transverse field (HTF)-$\mu$SR measurement, the curve-fit analysis is conducted using the equation:
\begin{align}
  A(t)=A_\alpha\exp\left(-\frac{\Delta^2t^2}{2}\right)\cos\left(\omega t+\phi_\alpha\right),
\end{align}
where $A_\alpha$ ($\alpha=\pm\hat{x},\pm\hat{y}$) is the asymmetry at $t=0$, and $\phi_\alpha$ is the initial phase of the spin rotation.
We note that the V and V' sites are not discernible by the frequency shift [as inferred from the fast Fourier transform (FFT) of the HTF-$\mu$SR spectra shown in Supplemental Material~\cite{supplement}, Fig.~S3]. Therefore, we adapt a single-component analysis in which the shift, $K_\mu$, is evaluated from the angular frequency, $\omega$, as follows:
\begin{align}
  \label{Eq2}
  K_\mu&=\dfrac{\omega-\gamma_\mu B_\mathrm{ref}}{\gamma_\mu B_\mathrm{ref}}-K_\mathrm{D},\\
  K_\mathrm{D}&=(4\pi/3-N_z)\chi,
\end{align}
where $\gamma_\mu/2\pi=135.53$ [MHz/T] is the muon gyromagnetic ratio; $K_\mathrm{D}$ (expressed in cgs units) is the correction term for the Lorentz and demagnetization fields; $N_z$ is the demagnetization factor, which depends on the geometry of the sample~\cite{Nz}; and $\chi$ is the uniform susceptibility.
The reference field, $B_\mathrm{ref}$, is simultaneously monitored through the precession frequency of the muons stopped in a scintillator slab made of a nonmagnetic insulator, CaCO$_3$, placed immediately behind the sample.
This scintillator serves as a muon ``veto'' counter so that the muons stopped in the sample and those stopped in the scintillator slab can be recorded separately.
The instrument-specific temperature drift in $B_\mathrm{ref}$ is calibrated through additional measurements on a silver sample.
\begin{figure}[!t]
  \centering
	\includegraphics[width=\linewidth]{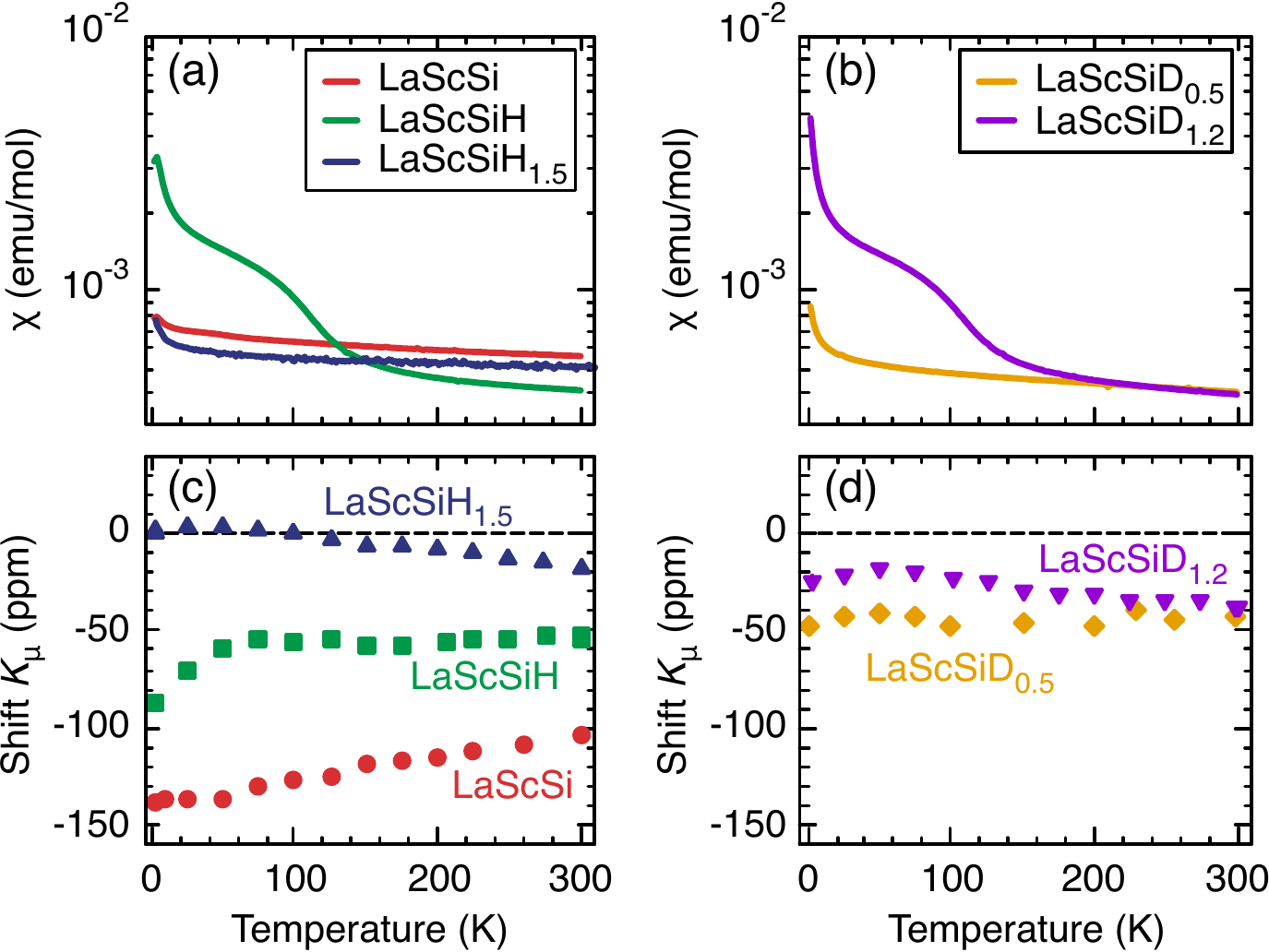}
  \caption{Temperature dependence of the uniform magnetic susceptibility, $\chi$ (a,b), and $\mu$SR frequency shift, $K_\mu$ (c,d), in LaScSiH/D$_x$. The error bars are smaller than the symbol size.} 
  \label{Fig_shift}
  \includegraphics[width=0.7\linewidth]{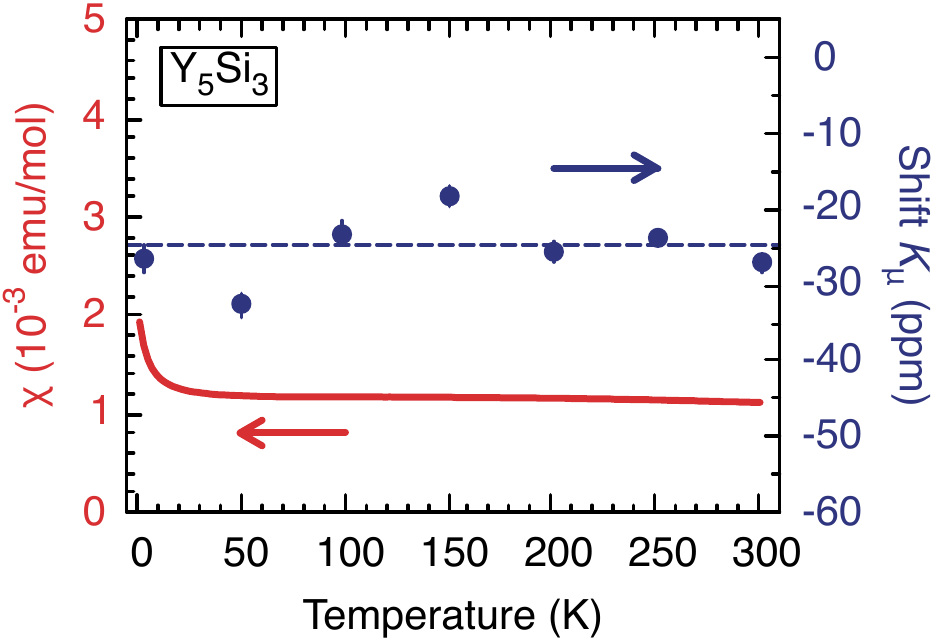}
  \caption{Temperature dependence of the uniform magnetic susceptibility, $\chi$ (left axis), and $\mu$SR frequency shift, $K_\mu$ (right axis), in Y$_5$Si$_3$. The dashed line represents the average value over temperature.} 
  \label{Fig_shift_Y5Si3}
\end{figure}

\par
The temperature dependence of $\chi$ and $K_\mu$ for LaScSi and its hydrides/deuterides is shown in Fig.~\ref{Fig_shift}.
It is noticeable that $\chi$ exhibits a significant increase below approximately 150~K in LaScSiH and LaScSiD$_{1.2}$, which is probably due to the unknown impurity phase(s).
In other samples, $\chi$ exhibits mostly temperature-independent (Pauli paramagnetism-like) behavior.
The magnitude of $\chi$ above approximately 150~K is the largest in pristine LaScSi, and this is in agreement with the results reported in the literature~\cite{WuAdvMat17}.
This is also consistent with the results of the DFT calculations, where the density of states near $E_\mathrm{F}$ is the largest in LaScSi.
We note that the demagnetization correction ($K_\mathrm{D}$) may be inaccurate at lower temperatures for LaScSiH and LaScSiD$_{1.2}$ because the relative uncertainty associated with $N_z$ is increased by the larger $\chi$ below approximately 150~K.
We can assume that $K_\mu$ is negative and mostly independent of temperature.
In particular, $K_\mu$ in LaScSi is as large as $-150$ ppm. This cannot be easily explained by the conventional chemical shifts as observed in C12A7~\cite{C12A7_muon}.
\par
The magnitude of $K_\mu$ in hydrogenated LaScSi exhibits a general decreasing trend with increasing H content. The apparently weaker dependence on $x$ in the deuterated samples may be attributed to the uncertainty in $x$ (see Sec.~\ref{SecII}). As H is mostly in hydride ions at the V/V' sites, the reduction in $K_\mu$ is uniquely ascribed to the decrease in the number of electride electrons due to their partial adsorption into H. This trend also suggests that the rest of the electride electrons remain delocalized while H intake occurs; $K_\mu$ will be unchanged if the Mu probes nearby localized electrons.
\par
Figure~\ref{Fig_shift_Y5Si3} shows the temperature dependence of $\chi$ and $K_\mu$ in Y$_5$Si$_3$.
Except for the small Curie term at temperatures below approximately 20~K, $\chi$ is almost independent of temperature, indicating that the apparent Pauli paramagnetic contribution dominates.
 $K_\mu$ also exhibits an almost temperature-independent negative shift of approximately $-25$~ppm, which is qualitatively similar to that in LaScSiH$_x$.
(The FFT spectra are shown in Supplemental Material~\cite{supplement}, Fig.~S3.)
\par
The results for $K_\mu$ in LaScSiH/D$_x$ and Y$_5$Si$_3$, together with the large negative shifts previously observed in single crystalline Y$_2$C~\cite{Y2C_muon}, provide compelling evidence that the anomalous diamagnetism inferred from $\mu$SR in these compounds is an important hallmark of electride electrons in intermetallic compounds involving transition metals.

\section{DISCUSSION}
In general, the uniform susceptibility of transition metal compounds is the sum of contributions from various sources:
\begin{align}
\chi = \chi_\mathrm{dia} + \chi_s + \chi_d + \chi_\mathrm{vv},
\end{align}
where $\chi_\mathrm{dia}$ is the diamagnetism of conduction electrons, $\chi_{s}$ ($\chi_{d}$) is the spin susceptibility of $s$ ($d$) band electrons, and $\chi_\mathrm{vv}$ is the Van Vleck orbital susceptibility.
The corresponding muon Knight shift is then expressed as
\begin{align}
K_\mu = K_\mathrm{dia} + K_s +K_d +K_\mathrm{vv}.
\end{align}
The sum of $\chi_s$ (Pauli paramagnetism) and $\chi_\mathrm{dia}$ ($=-\chi_s/3$ for Landau diamagnetism) gives a positive and small contribution to $K_\mu$ in simple metals, whereas the contributions of $d$ electrons can be large, with its signs determined by the hyperfine field ($H_\mathrm{hf}$).
In conventional metals, relatively large diamagnetic shifts have been observed by $\mu$SR only in metals with strong $s$-$d$ hybridization, such as Ni, Pd, and Pt, in which the contribution of $d$ electrons has been inferred from the Curie-Weiss behavior of $\chi_d$ and $K_d$~\cite{Schenk_shift,JASeitchik_Pd_NMR,Segransan_NMR_Ni,FNGygax_Ni,FNGygax_Pt}.
In LaScSi and Y$_5$Si$_3$ (as in Y$_2$C), strong hybridization of the electride electron and rare-earth atoms has been reported~\cite{Zhang_Y2C, Y5Si3,WuAdvMat17}.
However, the absence of the Curie-Weiss behavior in $K_\mu$ suggests that no such contribution exists in the electrides.
\par
One possible origin of the temperature-independent diamagnetism is the Van Vleck term associated with the electride electrons:
\begin{align}
\chi_\mathrm{vv}=2N\sum_i\frac{|\langle i|J_z|0\rangle|^2}{E_i-E_0},
\end{align}
where $N$ is the density of atoms per unit lattice, $J_z$ is the angular momentum operator, $|i\rangle$ is the excited state that has a finite orbital angular momentum, $|0\rangle$ is the ground state, and $E_i$ and $E_0$ are their respective eigenenergies.
Here, the contribution of the $d$ orbitals is expected to be larger owing to the strong $s$-$d$ hybridization with the electride electrons.
Because $E_i-E_0$ is generally sufficiently larger than $K_\mathrm{B}T$, $\chi_\mathrm{vv}$ does not vary with temperature. The negative sign of $K_\mu$ can be reasonably attributed to the magnetic dipole character of $H_\mathrm{hf}$ between the Mu and the off-site $d$ electrons, where the shift is given by the following relation: $K_\mu=H_\mathrm{hf}\chi_\mathrm{vv}$.

Another possibility is an anomaly in the Landau diamagnetism due to the Dirac electron-like dispersion relation~\cite{Fuseya:15}. The DFT calculations for these electrides suggest a slight deviation from the quadratic dispersion relation for the electride bands near the Fermi level, which may lead to an enhanced $\chi_\mathrm{dia}$ (not exceeding Pauli paramagnetism) and $K_\mathrm{dia}$~\cite{Hamada}.

A closer look at Fig.~\ref{Fig_shift}(c,d) shows that $K_\mu$ exhibits a slight decrease with increasing temperature in LaScSi. As $x$ increases, $K_\mu$ at 300~K approaches zero. The temperature-dependent trend in LaScSiH$_{1.5}$ is opposite that in LaScSi.
This trend can be explained by the temperature variation of $f_\mathrm{V}$, which leads to a change in $|K_\mu|$ as the average of the different $K_\mu$ values at the V and V' sites.
\par
Finally, the absence of such a large diamagnetism in the prototype electride, C12A7:e$^-$, is probably due to the absence of overlaps with other orbitals~\cite{c12a7_matsuishi2008,C12A7_muon}.
The relatively small occupancy of the electride electron sites (1/3 or less) and the associated small density near the Mu may also contribute to the negligible diamagnetism.


\begin{acknowledgments}
We would like to thank the TRIUMF staff for their technical support during the $\mu$SR experiment and N. Hamada for stimulating discussion.
This work was supported by the MEXT Elements Strategy Initiative to Form Core Research Center for Electron Materials (Grant No. JPMXP0112101001) and JSPS KAKENHI (Grant Nos. 19K15033 and 17H06153).
The $\mu$SR experiments were conducted under user programs (Proposal No. 2013MS01) at the Materials and Life Science Experimental Facility of the J-PARC and M1437 at TRIUMF.
The magnetic susceptibility measurements were performed using the instrument at the Neutron Science and Technology Center, CROSS.
\end{acknowledgments}

%

\end{document}


\title{{\Large Supplemental Material}\\
  \vspace{3mm}
  for\\
  \vspace{3mm}
  Anomalous diamagnetism of electride electrons in transition metal silicides}

\author{M.~Hiraishi}
\affiliation{\IMSS}
\author{K.~M.~Kojima}
\affiliation{\IMSS}\affiliation{\Sokendai}
\author{H.~Okabe}
\affiliation{\IMSS}
\author{A.~Koda}
\affiliation{\IMSS}\affiliation{\Sokendai}
\author{R.~Kadono}
\affiliation{\IMSS}\affiliation{\Sokendai}
\author{J.~Wu}
\affiliation{\MCES}
\author{Y.~Lu}\affiliation{\MCES}
\author{H.~Hosono}
\affiliation{\MCES}

\maketitle
\subsection{1. Experimental Methods and DFT calculations}

Conventional $\mu$SR measurements were performed using the ARTEMIS spectrometer installed in the S1 area of the Material and Life Science Experimental Facility, J-PARC in Japan~\cite{ARTEMIS}. [These measurements were omitted for Y$_5$Si$_3$ because of the negligible depolarization rate (approximately 0.0125~$\mu$s) due to the small nuclear magnetic moments].
The time evolution of the muon spin polarization ($\mu$SR time spectrum) was monitored through the decay positron asymmetry, $A(t)$, to obtain the local distribution of the internal magnetic field under zero and weak longitudinal magnetic fields.
For frequency shift measurements, $\mu$SR experiments were conducted under a high transverse magnetic field of 6~T using the NuTime spectrometer installed on the M15 beamline of TRIUMF, Canada.
\par
The uniform magnetic susceptibility, $\chi$, was measured using a Quantum Design SQUID magnetometer from 2 to 300~K under a magnetic field of 1~T.
\par
DFT calculations based on the generalized gradient approximation and the norm-conserving pseudopotential method were performed using the OpenMX code package~\cite{openmx}.
The cutoff energy was 250~Ry, and the structure was relaxed until the maximum force acting on each atom was less than 1$\times10^{-4}$~Hartree/Bohr.
A GGA-PBE exchange-correlation functional was used together with a $12\times12\times3$ mesh of $K$ points in the unit cell.

\subsection{2. Relaxation rate $\Delta$ in Kubo-Toyabe function}
The linewidth $\Delta$ is evaluated as the sum of contributions from the $m$-th type of nuclear magnetic moments, where $m = 1,2,$ and 3 for $^{139}$La, $^{45}$Sc, and $^{29}$Si, respectively.
\begin{align}
  \Delta_i^2&\simeq\gamma_\mu^2\sum_{m}f_m\sum_j\sum_{\alpha=x,y}\sum_{\beta=x,y,z}\gamma_m^2({\bf \hat{A}}_{mj}{\bf I}_m)^2 \label{dlts}\\
  {\bf \hat{A}}_{mj}&=A^{\alpha\beta}_{mj}=(3r_{mj}^\alpha r_{mj}^\beta-\delta_{\alpha\beta}r_{mj}^2)/r_{mj}^5\quad(\alpha, \beta=x,y,z)\nonumber
\end{align}
Here, $\gamma_\mu/2\pi=135.53$ [MHz/T] is the muon gyromagnetic ratio, ${\bm r}_{mj}=(x_{mj},y_{mj},z_{mj})$ is the position vector of the $j$-th nucleus from the Mu at the $i$-th site, and ${\bm \mu}_m=\gamma_m{\bf I}_m$ is the nuclear magnetic moment, which has the natural abundance of $f_m$. Because $^{139}$La and $^{45}$Sc have spin $I_m\ge1$, ${\bm \mu}_m$ is subject to electric quadrupolar interactions with the electric field gradient generated by the point charge of the diamagnetic Mu. This leads to a reduction in the effective ${\bm \mu}_m$ (by a factor $\sqrt{2/3}$ in the classical limit) to the value parallel to ${\bm r}_{mj}$.

\subsection{3. Temperature dependence of parameters in ZF/wLF data}
\begin{figure}[htbp]
  \centering
  \includegraphics[width=\linewidth]{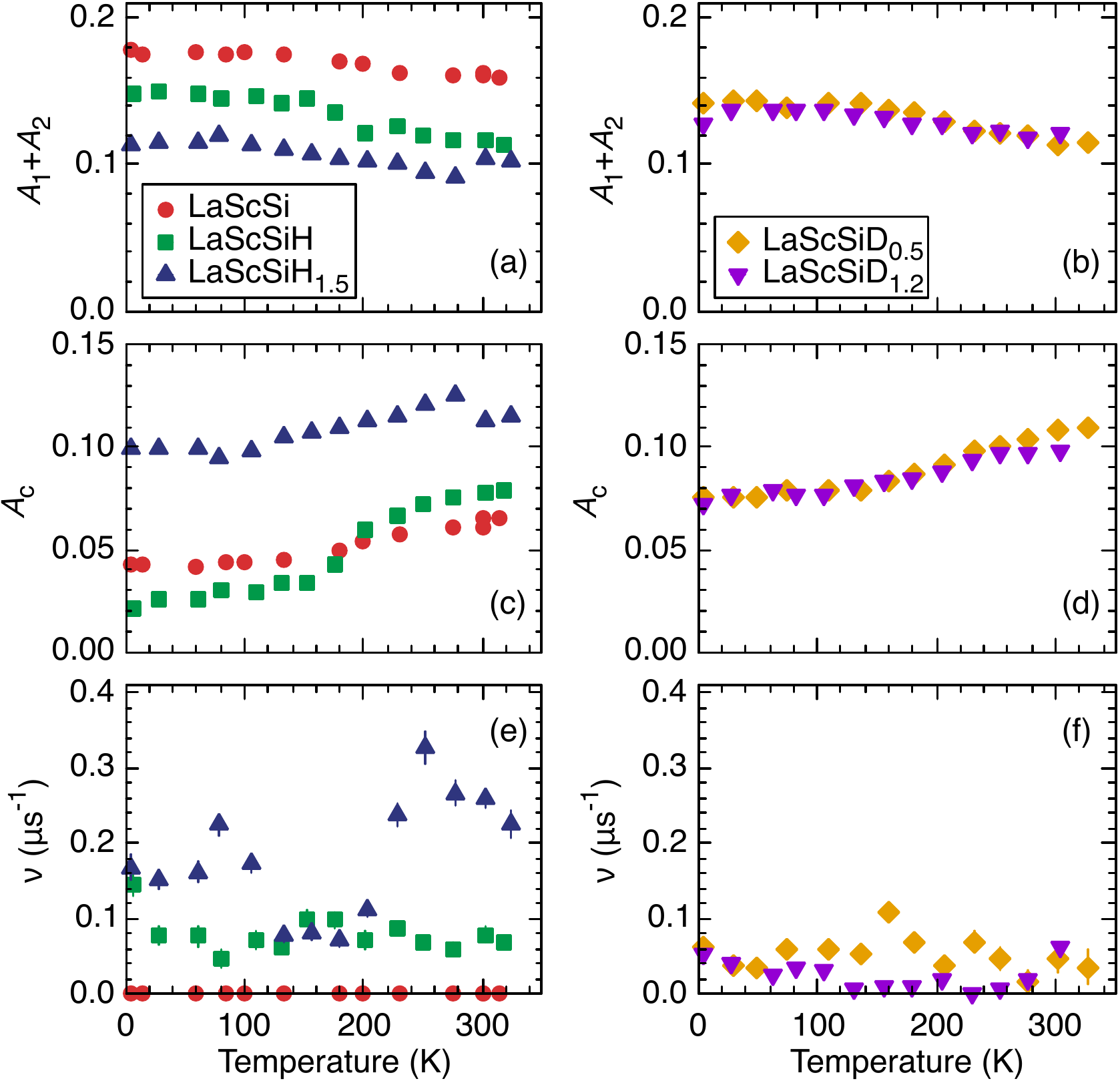}
  \caption{Temperature dependence of $A_1+A_2$ (a,~b), $A_\mathrm{c}$ (c,~d) and $\nu$ (e,~f) obtained by Eq.~(1). The figure on the left (light) shows LaScSi and its hydrogenated (deuterated) samples.}
  \label{FigS_asy_nu}
\end{figure}

The parameters obtained from the simultaneous analysis of the zero field (ZF) and weak longitudinal field (LF)-$\mu$SR for each sample are shown in Fig.~\ref{FigS_asy_nu}.
The function used in the curve-fit analysis is the sum of the dynamic Kubo-Toyabe function and a constant term.
\begin{align}
  A_0G_z(t)=\sum_{i=1}^nA_iG_\mathrm{KT}(\Delta_i,B_\mathrm{ext},\nu_i,t)+A_\mathrm{c}
  \label{Eq1}
\end{align}
We analyzed the data using several patterns (e.g., $n=1$ or 2, and $\nu_i$ is fixed at 0 or finite value) and adopted the pattern that minimizes the $\chi^2$ value in the least-squares method with as few parameters as possible.
LaScSi was analyzed $n=2$ and $\nu_i$ fixed at 0, while LaScSiH and LaScSiH$_{1.5}$ were analyzed with $n=2$, $\nu_1$ fixed at 0 and $\nu_2$ as the fit parameter.
LaScSiD$_{0.5}$ was analyzed with $n=2$ and $\nu_1$ fixed as the fit parameter and $\nu_2$ fixed at 0.
LaScSiD$_{1.2}$ was analyzed with $n=1$ and $\nu_1$ was set as the fit parameter.
Due to differences in sample size and other factors, the contribution from non-samples (e.g., sample holders) to $A_\mathrm{c}$ is considered to be the largest for LaScSiH$_{1.5}$, followed by deuterated samples, and the smallest for LaScSiH and LaScSi, resulting in $A_1+A_2$ and $A _\mathrm{c}$ are not directly comparable between different samples.

\subsection{4. A model for the H/Mu site occupancy of the asymmetric double-well potential}
We consider a simple model for the temperature dependence of the H/Mu site occupancy in the presence of two sites with asymmetric double-well potential separated by a potential barrier. Provided that the energy at the V' site is greater than that at the V site by $\varDelta\epsilon$, the partition function for the two-level system is given by
\begin{align}
  Z(\beta)=n+n'e^{-\beta\varDelta\varepsilon},
\end{align}
where $\beta\equiv 1/k_\mathrm{B}T$, $n$ and $n'$ are the degeneracy of the each site in the unit cell; $n=2$ and $n'=1$ for LaScSi. The fractional occupancy of H for the respective sites in the equilibrium state is then described by
\begin{align}
  f_\mathrm{V}=\frac{2}{2+e^{-\beta\varDelta\varepsilon}},\label{fv}\\
  f_\mathrm{V'}=\frac{e^{-\beta\varDelta\varepsilon}}{2+e^{-\beta\varDelta\varepsilon}}\label{fvv}.
\end{align}
Note that $f_\mathrm{V}<1$ at finite temperatures to reduce the free energy by gaining entropy.

Now, we presume that the initial site occupancy for Mu is that quenched from $T=\infty$ ($\beta=0$), so that $f_\mathrm{V}=2/3\equiv f^0_\mathrm{V}$, $f_\mathrm{V'}=1/3\equiv f^0_\mathrm{V'}$ (i.e., proportional to the number density of available sites). Then, the fractional yields observed by $\mu$SR at the finite temperature correspond to the average fraction of muons over the annealing process from this initial distribution to the thermal equilibrium distribution (Eqs.~\ref{fv},~\ref{fvv}) in the time scale of $\sim$$10^1$ $\mu$s. Such a relaxation process is generally described by the fluctuation-dissipation theorem within the linear response theory for the macroscopic systems.

\begin{figure}[htbp]
  \centering
  \includegraphics[width=0.9\linewidth]{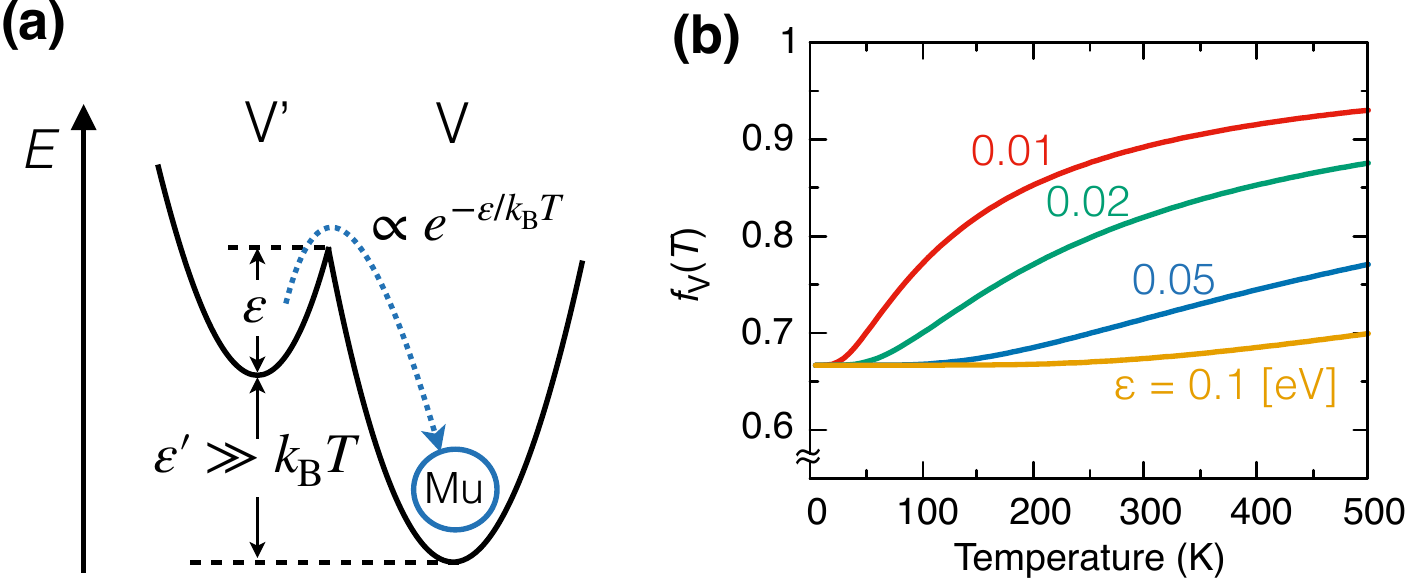}
  \caption{(a): Schematic drawing of the two-site model based on the Boltzmann factor. (b): Temperature dependence of Eq.~(\ref{Eq_twosite}) for various $\varepsilon$ in $x=0$ ($f_\mathrm{V}^0=2/3, f_\mathrm{V'}^0=1/3$).}
  \label{FigS_twosite}
\end{figure}

However, since the implanted Mu as microscopic entity probes the local fluctuations only, we assume that the observed temperature dependence of $f_\mathrm{V}$ is determined by the migration from the V' to V sites via thermally activated hopping over a potential barrier $\varepsilon$ [see Fig.~\ref{FigS_twosite}~(a)], where the migration probability is proportional to $e^{-\beta\varepsilon}$. The observed fraction of muons at the V' site can be approximately given by $f_\mathrm{V'}(T)=f_\mathrm{V'}^0\left(1-e^{-\beta\varepsilon}\right)$, which is valid for low temperatures where the inverse hopping process is negligible. Considering the relation $f_\mathrm{V}(T)+f_\mathrm{V'}(T)=1$, we have
\begin{align}
  f_\mathrm{V}(T)&=1-f_\mathrm{V'}(T)\nonumber \\
  &=1-f_\mathrm{V'}^0\left(1-e^{-\beta\varepsilon}\right)\nonumber \\
  &=f_\mathrm{V}^0+(1-f_\mathrm{V}^0)e^{-\beta\varepsilon},
  \label{Eq_twosite}
\end{align}
for the temperature dependence of the Mu occupancy at the V site.
Fig.~\ref{FigS_twosite}~(b) shows $f_\mathrm{V}$ given by Eq.~(\ref{Eq_twosite}) for various $\varepsilon$ in the sample with $x=0$. We analyzed the temperature dependence of the Mu site occupancy with $f_\mathrm{V}^0$ and $\varepsilon$ as fitting parameters.

\subsection{5. Fast Fourier transforms of HTF data}
\begin{figure*}[htbp]
  \centering
  \includegraphics[width=0.7\linewidth]{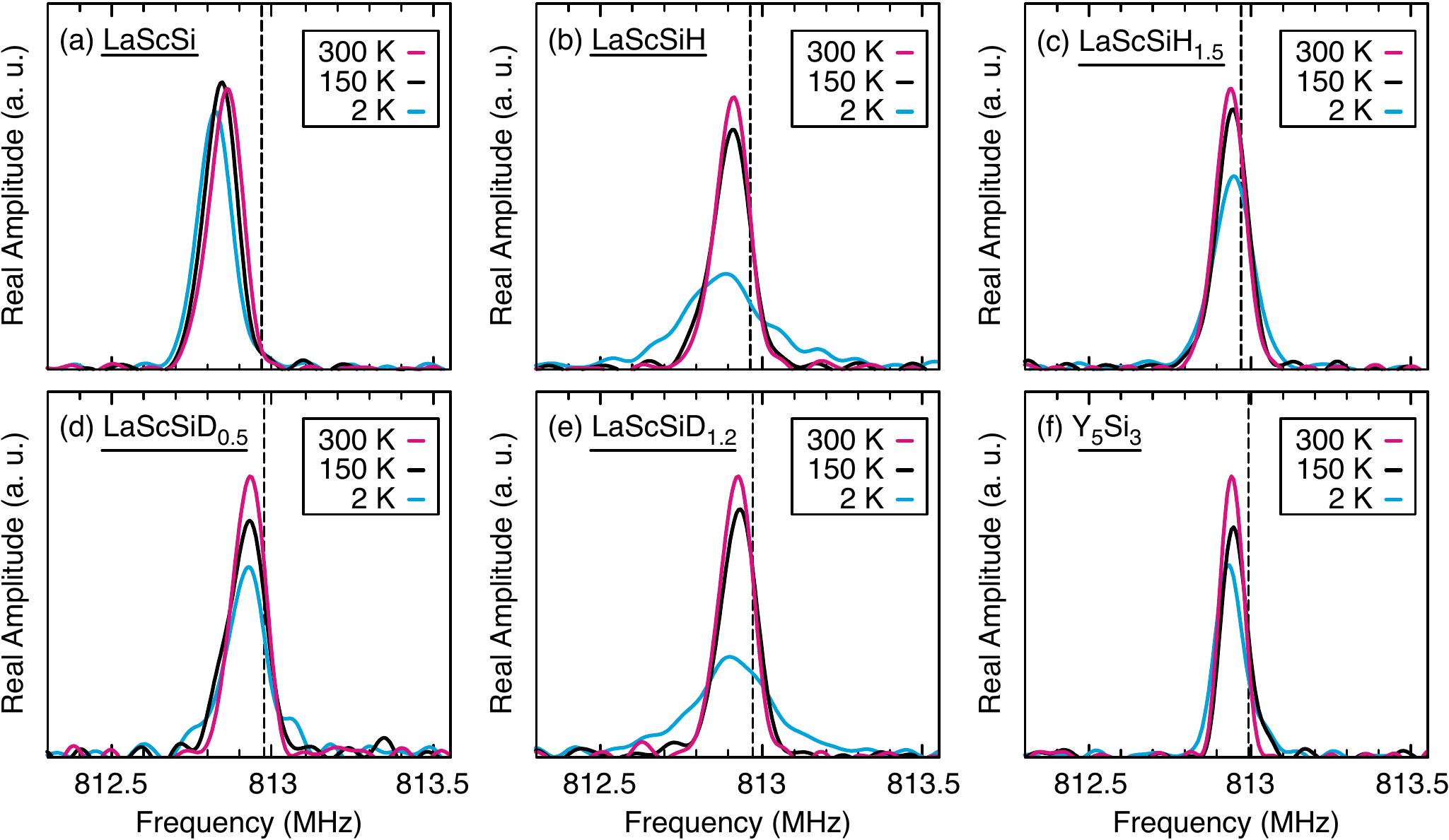}
  \caption{Fast Fourier transformed $\mu$SR spectra observed in (a) LaScSi,  (b,c) hydrogenated samples,  (d,e) deuterated samples, and  (f) Y$_5$Si$_3$ under a transverse field of 6~T. The vertical dashed lines are the precession frequency of CaCO$_3$ at 300~K representing the reference frequency of the shift ($K_\mu=0$).}
  \label{FigS_fft}
\end{figure*}

The fast Fourier transforms (FFTs) of the time spectra obtained under an external transverse field of 6~T are shown in Fig.~\ref{FigS_fft}.
The FFTs of LaScSiH (b) and LaScSiD$_{1.2}$ (e) at 2~K show a broad distribution, which may be due to the significant increase in the uniform susceptibility $\chi$ appearing below 150~K [see Fig.~4(a,~b) in the main text].

